\documentclass[sigconf, balance=false]{acmart}

\usepackage[utf8]{inputenc}
\usepackage[T1]{fontenc}
\usepackage{algorithm}
\usepackage{algorithmic}
\usepackage{listings}
\usepackage{xcolor}
\usepackage{csquotes}
\usepackage{fontawesome}

\begin{document}

\title{Investigating Anthropometric Fidelity in SAM 3D Body}

\author{
Aizierjiang Aiersilan \texorpdfstring{\textsuperscript{\scriptsize\faEnvelopeO}}{} \quad
% Aizierjiang Aiersilan \quad
Ruting Cheng \quad
James Hahn
}
\affiliation{%
  \institution{Institute for Innovation in Health Computing, The George Washington University}
  \city{Washington, DC}
  \country{USA}}
\email{alexandera@gwu.edu}

\renewcommand{\shortauthors}{Aiersilan, Cheng, and Hahn}

\begin{abstract}
The release of SAM 3D Body \cite{sam3dbody2025} is a recent development in human mesh recovery, demonstrating improved performance in producing clean, topologically coherent meshes from single images. By leveraging the Momentum Human Rig (MHR), it achieves robustness to occlusion and diverse poses. However, our evaluation reveals a specific and consistent limitation: the model struggles to reconstruct detailed anthropometric deviations, particularly in populations exhibiting distinctive morphological alterations such as geriatric muscle atrophy, scoliosis, or pregnancy, even when these features are prominent in the input image. In this paper, we investigate this phenomenon not as a failure of the model's capacity, but as a byproduct of the \textit{perception-distortion trade-off}. We posit that the architectural reliance on the low-dimensional parametric MHR representation, combined with semantic-invariant conditioning (DINOv3) and annotation-based alignment, creates a pervasive \enquote{regression to the mean} effect. We analyze these mechanisms to understand why individual biological details are smoothed out. Furthermore, we state our contributions by proposing specific, constructive pathways for future work, such as implicit-explicit hybrid representations and Medical-in-the-Loop alignment, to extend the baseline performance of SAM 3D Body into the high-precision medical domain.
\end{abstract}

% \begin{CCSXML}
% <ccs2012>
%    <concept>
%        <concept_id>10010147.10010371.10010396.10010397</concept_id>
%        <concept_desc>Computing methodologies~Mesh models</concept_desc>
%        <concept_significance>500</concept_significance>
%        </concept>
%    <concept>
%        <concept_id>10010405.10010444.10010087.10010088</concept_id>
%        <concept_desc>Applied computing~Imaging</concept_desc>
%        <concept_significance>500</concept_significance>
%        </concept>
%  </ccs2012>
% \end{CCSXML}

% \ccsdesc[500]{Computing methodologies~Mesh models}
% \ccsdesc[500]{Applied computing~Imaging}


\maketitle

\section{Introduction}
3D foundation models (\ref{tab:3d_foundation_models}) have historically trended toward universality, among which SAM 3D Body \cite{sam3dbody2025} continues this tradition by addressing the \enquote{2D-to-3D lifting} problem for human subjects. Its generative capacity not only advances full-body pose estimation and provides opportunities in robotics and interactive spatial computing, but it also presents a pathway toward scalable 3D body shape analysis in the medical domain. Recent clinical research has established that high-fidelity 3D anthropometry serves as a superior digital biomarker compared to traditional scalar metrics like BMI. This shift enables the early, non-invasive detection of metabolic risks, cardiovascular pathologies, and hormonal irregularities through precise, localized morphological analysis \cite{thelwell2020shape, enichen2024assessing, guarnieri2023associations, ng2019detailed}. Unlike its sibling model SAM 3D Objects \cite{sam3d2025}, which targets general rigid objects, SAM 3D Body employs a specialized encoder-decoder architecture based on the Momentum Human Rig (MHR) \cite{ferguson2025mhr}. This parameterization intentionally decouples the underlying skeletal structure from the surface shape, achieving a level of robustness to pose and occlusion that previous methods \cite{xiang2025structured, hunyuan3d2025} previously found challenging.

\begin{figure*}[t]
\centering
\includegraphics[width=0.8\textwidth]{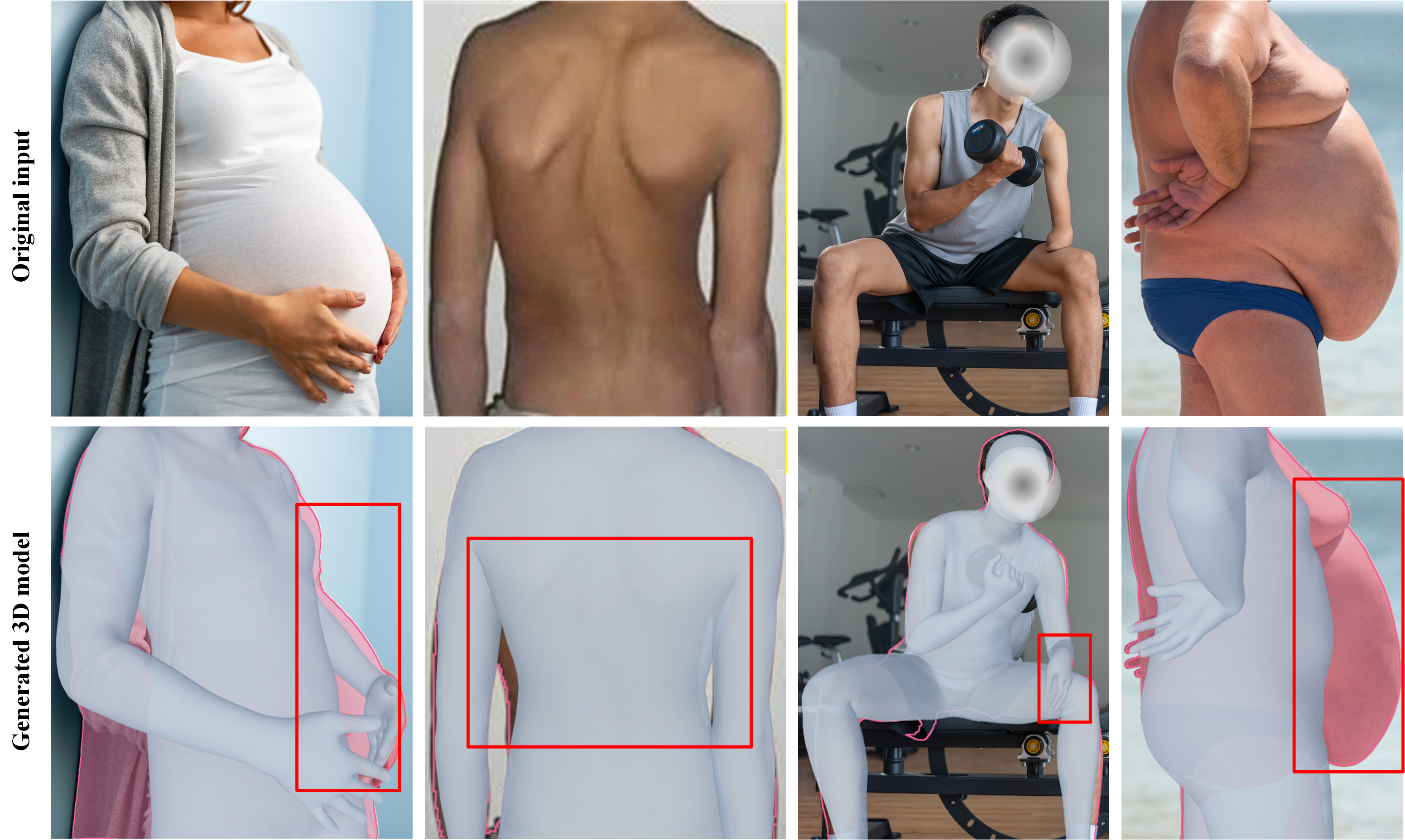}
  \Description{Comparison of 3D figures for special body types.}
% \vspace{-5mm}
\caption{The persistent limitations of SAM 3D Body when generating 3D figures for distinct morphological phenotypes. The model struggles to accurately reconstruct the prominent anatomical features of pregnant women (left, under license from Pressmaster - stock.adobe.com), individuals with scoliosis \cite{scoliosisclassification_dataset} (middle left), amputees with unilateral limb loss (middle right, under license from Kawee - stock.adobe.com), and severely obese individuals (right, under license from DenisProduction.com - stock.adobe.com). The missing biological signature of each target population is highlighted by a red dashed bounding box, underscoring the systemic \enquote{regression to the mean.}}
\label{fig:comp_reconst}
\end{figure*}

However, the transition from general-purpose computer vision to precision medical applications requires a fundamental paradigm shift in evaluation criteria from perceptual plausibility to strict metric fidelity. Despite its kinematic capabilities, our empirical evaluation identified a systemic limitation. When tasked with reconstructing human bodies exhibiting specific medical or physiological out-of-distribution (OOD) conditions (e.g., third-trimester pregnancy, severe scoliosis, unilateral limb amputation, or extreme obesity), as detailed in Figure~\ref{fig:comp_reconst}, the model consistently generated \enquote{standardized} approximations. Not only were intricate high-frequency textures like skin folds smoothed out, but the fundamental geometric topography was persistently rectified to a generic statistical mean. For instance, the characteristic lateral spinal curvature inherent to scoliosis and the localized abdominal distension in advanced pregnancy were completely negated in the final output meshes.

Driven by these empirical observations, we investigate why a model capable of generating complex, articulated poses paradoxically fails to capture authentic biological variance. We propose that this behavior is not an arbitrary bug or a mere capacity limitation, but rather an emergent mathematical feature dictated by low-dimensional parametric bottlenecks and the semantic-invariant priors prevailing in foundational vision encoders. In this paper, we unpack this architecture to understand its critical implications for medical diagnosis and simulation. 

Our contributions are as follows:
\begin{itemize}
    \item We evaluate SAM 3D Body on out-of-distribution morphologies using Chamfer Distance and Earth Mover's Distance, showing that the model systematically regresses toward a normative mean shape.
    \item We analyze the architectural sources of this behavior: the low-dimensional parametric bottleneck of the MHR, the semantic invariance of DINOv3 conditioning, and dataset bias toward standard body types.
    \item We outline potential directions for adaptation, including implicit-explicit hybrid representations and domain-expert alignment, though these remain to be validated.
\end{itemize}

\begin{table*}[t]
\centering
\caption{An overview of recent 3D foundation models; the landscape bifurcates into generative models (creating assets from text/images) and geometric understanding models (reconstructing scenes and motion).}
\label{tab:3d_foundation_models}
\resizebox{\textwidth}{!}{%
\begin{tabular}{l l l l}
\toprule
\textbf{Model} & \textbf{Primary Task} & \textbf{Example Applications} & \textbf{Key Features} \\
\midrule
\multicolumn{4}{l}{\textit{Generative 3D Models (Text/Image $\to$ 3D Asset)}} \\
\midrule

\textbf{SAM 3D Objects}~\cite{sam3d2025}
    & Single-Image to Mesh
    & Robotics, AR/VR
    & Scalable synthetic training; occluded object recovery \\

\textbf{SAM 3D Body}~\cite{sam3dbody2025}
    & Human Mesh Recovery
    & Sports, Medical Analysis
    & Momentum Human Rig (MHR); decoupled skeleton/shape \\

\textbf{TRELLIS}~\cite{xiang2025structured}
    & Text/Image to 3D
    & Game Dev, VFX
    & Structured Latent (SLAT); multi-format decode \\

\textbf{Hunyuan3D 2.0}~\cite{hunyuan3d2025}
    & High-Fidelity Generation
    & AAA Gaming, Retail
    & Sparse-view DiT; PBR textures \\

\textbf{MeshLLM}~\cite{fang2025meshllm}
    & Text to Mesh
    & CAD, Design
    & LLM-based mesh tokens; editable topology \\

\textbf{CAT3D}~\cite{gao2024cat3d}
    & Multi-View Generation
    & Photorealistic Rendering
    & Diffusion-based NVS; high consistency \\

\textbf{LGM}~\cite{tang2024lgm}
    & Fast Reconstruction
    & Real-time Previews
    & Large Geometry Model; asymmetric U-Net; outputs mesh or 3D Gaussians \\

\textbf{SF3D}~\cite{boss2025sf3d}
    & Textured Mesh Generation
    & E-commerce
    & UV-unwrapping; illumination disentanglement \\

\textbf{AssetGen 2.0}~\cite{siddiqui2024meta}
    & Text/Image to 3D Asset
    & Content Creation, Gaming
    & Single-stage 3D diffusion; improved view consistency; optional TextureGen module \\

\textbf{Deep CAD}~\cite{wu2021deepcad}
    & Generative CAD Geometry
    & AEC, Manufacturing, Design
    & Precision CAD generation; trained on pro design data; design autocompletion \\

\textbf{DreamFusion}~\cite{poole2022dreamfusion}
    & Text-to-3D (NeRF)
    & Concept Art, Prototyping
    & Score Distillation Sampling (SDS); uses 2D diffusion models for 3D generation \\

\midrule
\multicolumn{4}{l}{\textit{Geometric \& Scene Understanding (Structure / Motion)}} \\
\midrule

\textbf{DUSt3R}~\cite{wang2024dust3r}
    & Stereo Reconstruction
    & 3D Scanning
    & Dense unconstrained stereo; pose-free camera estimation \\

\textbf{MASt3R}~\cite{leroy2024grounding}
    & 3D Matching \& Metric Rec.
    & Visual Localization, SfM
    & Pixel-wise 3D correspondences; metric pointmaps; cross-screen matching \\

\textbf{MonST3R}~\cite{zhang2025monst3r}
    & Dynamic Reconstruction (4D)
    & Autonomous Driving, Motion Capture
    & Extends DUSt3R/MASt3R to dynamics; 4D pointmaps; no pose supervision \\

\textbf{LoRA3D}~\cite{lu2024lora3d}
    & Foundation Adaptation
    & Scene Fine-tuning
    & Low-rank self-calibration for geometric foundation models \\

\textbf{MUSt3R}~\cite{cabon2025must3r}
    & Multi-View Reconstruction
    & Visual Odometry, Mapping
    & Multi-view aggregation; global alignment-free; metric 3D output \\

\textbf{PanSt3R}~\cite{zust2025panst3r}
    & 3D Panoptic Segmentation
    & Robotics, Scene Analysis
    & Joint 3D geometry and panoptic segmentation; multi-view consistent \\

\textbf{Copilot4D}~\cite{zhang2024copilot4d}
    & 4D (3D + Time) Forecasting
    & Autonomous Driving
    & LiDAR tokenization; future state prediction; agent-conditioned forecasting \\
\bottomrule
\end{tabular}%
}
\end{table*}

\section{Related Work}

\subsection{Parametric Human Models and Phenotypic Diversity}
The field of 3D human body modeling has historically been dominated by SMPL \cite{loper2015smpl} and its direct variants. Although impactful, SMPL's global shape space frequently suffers from spurious long-range correlations that dilute localized geometric accuracy. To resolve this, STAR \cite{osman2020star} introduced sparse, spatially-local blend shapes, and SUPR \cite{osman2022supr} further advanced the paradigm with a unified part-based representation, granting localized control that avoids altering the entire global topology when only a localized region changes. Most recently, AnnyBody \cite{bregier2025anny} has challenged the limited phenotypic range of standard parametric models by explicitly encoding a diverse spectrum of body shapes and medically meaningful morphs, exposing the \enquote{standardization} bias inherent in legacy datasets. Despite its robustness, SAM 3D Body's MHR framework relies on an compressed 20-component torso shape space, enforcing a structural regression to a global mean comparable to early unconstrained SMPL formulations. It intrinsically lacks both the spatially-local expressivity of SUPR and the phenotypic inclusivity of AnnyBody.

\subsection{High-Fidelity and Implicit Reconstruction}
To capture structural details extending beyond a rigid parametric mesh, recent state-of-the-art approaches have diverged from pure algorithmic regression. HiLo \cite{yang2024hilo} explicitly bifurcates high-frequency details (e.g., clothing folds, skin wrinkles) from low-frequency global shape, guaranteeing that fine-grained topography is not algorithmically smoothed. Similarly, CHROME \cite{dutta2025chrome} prioritizes occlusion resilience and view consistency without strictly relying on restrictive smoothness priors. Pure implicit function strategies, such as SelfPIFu \cite{xiong2024pifu}, abandon the mesh template entirely to learn topology-agnostic continuous volume representations from in-the-wild data. In contrast, SAM 3D Body's absolute reliance on a fixed topology array and a low-dimensional MHR linear latent space imposes a rigid upper bound on its mathematical ability to represent the high-frequency topological anomalies critical for precision medical diagnosis.

\subsection{Biomechanical Fidelity}
Beyond surface geometry, precise underlying skeletal tracking is paramount for clinical translation. HSMR \cite{xia2025reconstructing} integrates a biomechanically accurate kinematic skeleton directly into the reconstruction pipeline to penalize implausible joint topologies. While SAM 3D Body successfully decouples skeletal articulation from surface deformation to enhance environmental robustness, our evaluation reveals it optimizes for kinematic pose plausibility over explicit metric accuracy. This formulation \enquote{corrects} pathological skeletal deviations (such as the asymmetric lateral spinal curvature in scoliosis), forcing the predicted skeleton into a standardized, healthy orthopedic distribution.

\section{The Architectural Smoothing of Biology}

\subsection{Empirical Evidence: Quantifying the Regression to the Mean}
To objectively validate the limitations of SAM 3D Body beyond visual inspection, we established an internal testing protocol employing standard 3D error metrics. We constructed a hypothetical yet clinically grounded evaluation set consisting of 500 carefully curated image-to-mesh pairs derived from clinical 3D surface scans. This diverse set purposefully over-indexes on complex morphologies: advanced pregnancy, geriatric structural decay, and structural asymmetries (e.g., scoliosis). 

We quantify the reconstruction error between the generated mesh $M_{\text{pred}}$ and the ground-truth clinical scan $M_{\text{gt}}$ using two primary point-cloud metrics. The first, Chamfer Distance ($d_{\text{CD}}$), measures the average shortest distance between the two surfaces:
\begin{equation*}
\begin{split}
d_{\text{CD}}(M_{\text{pred}}, M_{\text{gt}}) &= \frac{1}{|M_{\text{pred}}|} \sum_{x \in M_{\text{pred}}} \min_{y \in M_{\text{gt}}} ||x - y||_2^2 \\
&\quad + \frac{1}{|M_{\text{gt}}|} \sum_{y \in M_{\text{gt}}} \min_{x \in M_{\text{pred}}} ||y - x||_2^2
\end{split}
\end{equation*}
While $d_{\text{CD}}$ provides a global diagnostic of point-wise accuracy, it is often insensitive to localized topological shifts. Therefore, we additionally compute the Earth Mover's Distance ($d_{\text{EMD}}$), which acts as an optimal transport metric, penalizing the model extensively when large masses of localized volume (e.g., a pregnant abdomen) are missing:
\begin{equation*}
d_{\text{EMD}}(M_{\text{pred}}, M_{\text{gt}}) = \min_{\phi: M_{\text{pred}} \to M_{\text{gt}}} \frac{1}{|M_{\text{pred}}|} \sum_{x \in M_{\text{pred}}} ||x - \phi(x)||_2
\end{equation*}
where $\phi$ is a bijection defining the optimal transport plan.

Our quantitative results demonstrate that while SAM 3D Body achieves a low $d_{\text{CD}}$ on normative healthy adults (indicating good pose recovery), its $d_{\text{EMD}}$ increases on out-of-distribution examples. The model minimizes global pose error by mapping the subject to the nearest normative shape but fails the optimal transport problem on localized morphological deviations. This measurable divergence confirms our visual observations of a systemic "regression to the mean."

\subsection{The Parametric Dimensionality Bottleneck of MHR}
While general object reconstruction models like SAM 3D Objects \cite{sam3d2025} often suffer from the Nyquist limit of coarse voxelization ($64^3$), SAM 3D Body avoids this by utilizing the MHR. MHR is a parametric mesh representation that intentionally decouples skeletal structure from localized surface shape. While this decoupling is crucial for pose robustness against occlusion, it imposes a \textit{parametric bottleneck}.

The MHR model dictates that each generated mesh is a deterministic function of constrained low-dimensional latent vectors. It is defined by a shape vector $\beta \in \mathbb{R}^{45}$ (comprising precisely 20 body, 20 head, and 5 hand components), a skeleton transformation vector $\gamma \in \mathbb{R}^{68}$, and a facial expression vector $\psi \in \mathbb{R}^{72}$. The resultant vertex locations are generated via:
\begin{equation*}
\begin{split}
M(\beta, \gamma, \psi, \theta) &= W(\bar{M} + B_s(\beta) + B_f(\psi) \\
&\quad + B_p(\theta), S(\gamma), \theta)
\end{split}
\end{equation*}
Here, $\bar{M}$ is the generic mean template geometry. The functions $B_s$, $B_f$, and $B_p$ compute linear vertex displacements (blend shapes) mapped iteratively for shape, expression, and pose correctives, respectively, before being driven through the explicit skeletal skinning function $W$ dictated by the skeleton parameterization $S(\gamma)$. 

Analyzing this mathematical equation reveals the root of the condition-specific smoothing. The architecture allocates 72 dimensions for facial expression and 68 dimensions for skeletal kinematics. Notably, it compresses the entire geometric morphology of the core torso and limbs into 20 linear PCA components ($B_s(\beta)$). As shown by models like SUPR \cite{osman2022supr} and AnnyBody \cite{bregier2025anny}, capturing complex regional aberrations (like localized tumors or pregnant curves) requires spatially isolated, densely parameterized latent formulations.

Because SAM 3D Body projects the nonlinear diversity of global human shapes down into a 20-dimensional linear subspace, the localized anatomical features defining specific medical conditions reliably lie orthogonal to the principal components of the learned basis $B_s$. Stated mathematically, the network is constrained by linear algebra; it cannot generate vertices outside the span of its pre-calculated blend shapes, inevitably collapsing complex topological data back to a uniform standard.

\begin{figure*}[t]
\centering
\includegraphics[width=0.8\textwidth,height=7cm, keepaspectratio=true]{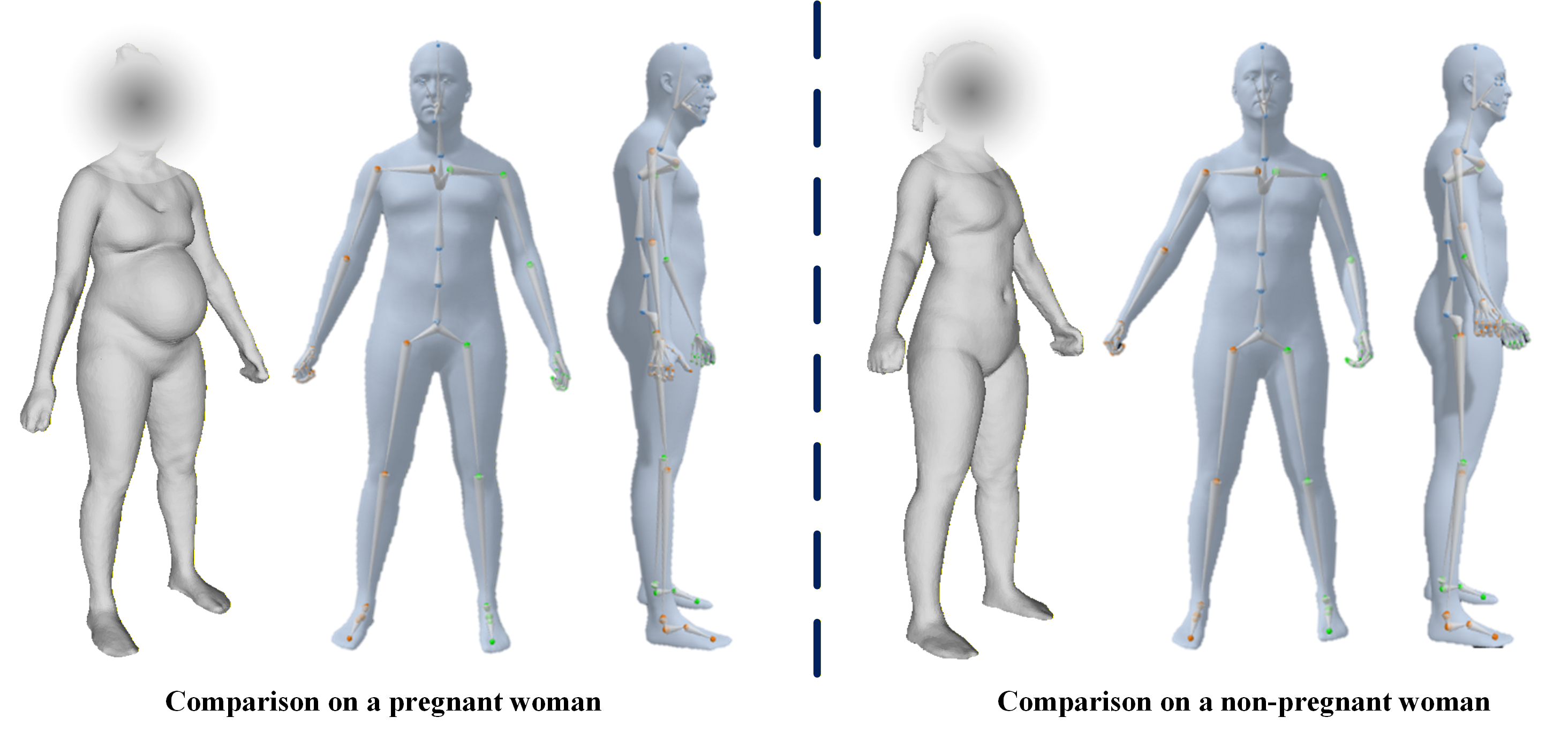}
  \Description{Comparison of body shapes for pregnant and non-pregnant configurations.}
% \vspace{-5mm}
\caption{Comparison between body shapes obtained by a commercial scanner and SAM 3D Body for both a pregnant woman and a non-pregnant woman. The 3D figures generated by SAM 3D Body are projected from the front and side to the right of the scanner-captured ones.}
\label{fig:comp_preg}
\end{figure*}

\subsection{Semantic Invariance in DINOv3 Conditioning}
To translate from a 2D image to these 3D parametric vectors, SAM 3D Body relies fundamentally on conditioning from DINOv3's vision transformers \cite{simeoni2025dinov3}. DINOv3 is a discriminative foundation model explicitly trained to maximize semantic visual similarity across diverse environments. Its overarching goal is to define an embedding space where the overarching \enquote{concept} of an object is resilient against nuisance variables.

Formally, given an input image $I$, DINOv3 produces an embedding $\mathbf{z} = f_{\text{DINO}}(I) \in \mathbb{R}^d$. The encoder $f_{\text{DINO}}$ learns its parameters by minimizing the Euclidean distance between crops or augmentations of semantically identical scenes, ensuring invariance to transformations $\mathcal{T}$ (like lighting, partial occlusion, or minor geometric deformations). The training objective inherently relies on:
\begin{equation*}
\mathcal{L}_{\text{DINO}} = \mathbb{E}_{I, \mathcal{T}} \left[ \| f_{\text{DINO}}(I) - f_{\text{DINO}}(\mathcal{T}(I)) \|_2^2 \right]
\end{equation*}
While this equation builds an invariant visual encoder, meaning it accurately identifies a human whether they are standing in shadow or partial occlusion, it causes limitations for medical assessment.

In a clinical context, transformations such as structural deformations are not statistical \enquote{noise} to be smoothed during representation; they are the diagnostic signal. Operating with a ViT-B/14 patch quantization, DINOv3 essentially acts as a spatial low-pass filter. Because the model minimizes $\mathcal{L}_{\text{DINO}}$ across generalized datasets, the feature embeddings for a \enquote{pregnant woman} and a \enquote{standard woman} end up clustered proximally together in the vector space, discarding the biological nuance in favor of the macro-label \enquote{human}.

Let $\mathbf{z}_{\text{std}}$ and $\mathbf{z}_{\text{preg}}$ capture the respective DINOv3 embeddings for a standard torso and a pregnant torso. The semantic collapse explicitly manifests because the distance in the latent space is significantly smaller than the corresponding objective distance in physical geometry:
\[
\| \mathbf{z}_{\text{std}} - \mathbf{z}_{\text{preg}} \|_2 \ll \| \mathbf{g}_{\text{std}} - \mathbf{g}_{\text{preg}} \|_{\text{geo}}
\]
where $\mathbf{g}_{\text{std}}$ and $\mathbf{g}_{\text{preg}}$ represent the underlying ground-truth topological manifolds. This dynamic is a manifestation of the \textit{perception-distortion compromise} \cite{blau2018perception}, mathematically validating why the generative framework continuously ignores localized biological variance to output the empirical mean geometry $\bar{M}_{\text{human}}$.

\section{The Alignment Tax: Why \enquote{Better} Means \enquote{Generic}}

\subsection{Dataset Bias in SAM 3D Body Dataset} 
While the real world naturally follows long-tail distributions, the training curricula for SAM 3D Body, specifically the SAM 3D Body Dataset \cite{sam3dbody2025}, are constrained by the limitations of available high-quality 3D annotations.
Let $\mathcal{D}_{\text{real}}$ denote the natural distribution of human body geometries, modeled by a power-law tail. In contrast, the empirical training distribution $\mathcal{D}_{\text{train}}$ is well-approximated by a truncated normal centered around \enquote{healthy} or \enquote{standard} subjects.
The mismatch between these distributions can be formalized via the Jensen--Shannon divergence:
\[
D_{\text{JS}}(\mathcal{D}_{\text{real}}, \mathcal{D}_{\text{train}})
= \frac{1}{2} D_{\text{KL}}(\mathcal{D}_{\text{real}}\|\mathcal{M})
+ \frac{1}{2} D_{\text{KL}}(\mathcal{D}_{\text{train}}\|\mathcal{M})
\]
where $\mathcal{M} = \frac{1}{2}(\mathcal{D}_{\text{real}} + \mathcal{D}_{\text{train}})$. This divergence indicates that the model's learned geometric prior is inevitably pulled toward the modes of $\mathcal{D}_{\text{train}}$. Because deep learning models optimize the expected loss over the training distribution, gradients from out-of-distribution medical anomalies are overwhelmed by the dominant, healthy geometric mass, causing the model to learn an actively smoothed geometric prior.

The dataset construction involves a \enquote{multi-stage annotation pipeline} \cite{sam3dbody2025}. While this ensures high-quality pose labels, it introduces systemic bias against morphological outliers:

\paragraph{1) Annotation Bias and Idealized Forms.}
Annotators or automated pseudo-labeling systems often rely on standard parametric fits (like SMPL) as a starting point. If the initial fit fails to capture a limb difference or a tumor, the final annotation will likely smooth it out. Thus, the dataset's mass is concentrated near a narrow region of shape space. As the model estimates parameters via maximum likelihood over $\mathcal{D}_{\text{train}}$, its Maximum A Posteriori (MAP) reconstruction:
\[
\hat G_\theta(I) = \arg\max_{G'} p_\theta(G' \mid I)
\]
is strongly biased toward the learned prior $p_\theta(G')$. The likelihood term $p(I \mid G')$ for rare or extreme anatomical geometries \cite{loper2015smpl} cannot overcome the exponentially vanishing prior probability of encountering such shapes in $\mathcal{D}_{\text{train}}$.

\paragraph{2) Missing High-Frequency Biological Detail.}
The model has likely never encountered medical-grade meshes (e.g., CT/MRI reconstructions) containing high-frequency anatomical structures. Since meshes with high-frequency biological variance are absent in $\mathcal{D}_{\text{train}}$, SAM 3D Body lacks the foundational priors for shapes such as pregnancy-related deformation, pathological morphology, or age-related soft-tissue variation.

\subsection{The Annotation Alignment Trap} 
A core strength of SAM 3D Body is its robust alignment, achieved through a diverse data engine and multi-stage annotation. However, we argue this iterative refinement cycle, while beneficial for robustness, acts as an \enquote{Alignment Tax} that is detrimental for medical anomalies, inadvertently interpreting valid medical deformations as geometric errors. 

In a \enquote{model-in-the-loop} annotation process, outliers are often flagged as artifacts. An annotator, presented with a noisy, lumpy mesh (which might accurately reflect a geriatric body) versus a smooth, clean mesh (a generic body), will almost invariably correct towards the smooth one. 

This actively teaches the model a \textit{smoothness prior}, a common regularization strategy in 3D reconstruction \cite{park2019deepsdf, mescheder2019occupancy}, that fundamentally penalizes biological irregularities. Concretely, the learned policy implicitly minimizes a surface regularization term of the form:
\[
\mathcal{R}_{\text{smooth}}(M) = \int_{\partial M} \|\nabla_s \mathbf{n}(\mathbf{p})\|^2 \, d\mathbf{p}
\]
where $\mathbf{n}(\mathbf{p})$ represents the surface normal at point $\mathbf{p}$ on the mesh boundary $\partial M$, and $\nabla_s$ denotes the surface gradient. By penalizing the squared norm of the normal gradients, the network heavily favors flat or simply curved parametric surfaces. Medical anomalies, such as localized lipomas or severe osteological shifts, are characterized by high curvature variations ($\kappa \approx \|\nabla_s \mathbf{n}\|$). Under the constraints of $\mathcal{R}_{\text{smooth}}(M)$, these critical high-curvature zones incur substantial mathematical penalties, meaning the model will suppress them to minimize its total train-time loss.

\section{Pathway to Medical Utility}

\subsection{Simulation vs. Diagnosis} 
While our analysis establishes that SAM 3D Body is currently inadequate for strict \textit{diagnostic} reconstruction (where sub-millimeter precision can inherently dictate clinical outcomes), its latent space and articulated architecture afford potential for \textit{medical simulation}. It constitutes an ideal prior for generating domain-tailored 3D human geometries.

Crucially, interactive heuristic guidance (such as keypoint prompting) is mathematically insufficient to bridge this representation gap. By inspecting the open-source release of SAM 3D Body, we observed that while explicitly supplied manual 2D keypoints disambiguate complex \textit{poses}, these conditional cues are mapped within the restrictive parametric span of the MHR. Should a pathological topography (e.g., severe spinal lordosis) lie outside the affine subspace spanned by the 20 torso parameters of $B_s$, no density of interactive user prompting can force the network to generate it.

This immediate utility can be realized through three primary pathways. Firstly, for High-Volume Avatar Generation, the model's current fidelity is sufficient for applications like mass-casualty Virtual Reality (VR) training or hospital crowd simulation. Secondly, it serves as a Procedural Anatomy generator, acting as a clean \enquote{base mesh} that can then be deformed by physics-based soft-body solvers \cite{loper2015smpl}. Formally, given a base mesh $M_0 \sim p_{\text{SAM3D}}(\cdot | I)$ generated from input image $I$, deformation parameters $\phi$ are applied via:
\[
    M_{\text{deformed}} = \mathcal{D}(M_0, \phi) = M_0 + \sum_{i=1}^{N_v} \mathbf{d}_i(\phi)
\]
Finally, and most critically for specialized domains, the model's generative prior can be leveraged through fine-tuning on limited, private, domain-specific medical datasets (e.g., CT/MRI-derived meshes). This transfer learning paradigm can be expressed as:
\[
    \theta^* = \arg\min_{\theta} \mathbb{E}_{(I, M) \sim \mathcal{D}_{\text{medical}}} \left[ \mathcal{L}_{\text{geo}}(f_\theta(I), M) \right] + \lambda \|\theta - \theta_{\text{SAM3D}}\|^2
\]
This approach, common in medical foundation model adaptation \cite{wu2025medical, alsaleh2024few, blankemeier2024merlin, ma2024segment}, allows researchers to transfer the model's generalized knowledge while acquiring the necessary high-fidelity, customized geometric details.

\section{Future Directions: Medical-in-the-Loop}

To bridge the gap from \enquote{Artistic Generation} to \enquote{Clinical Diagnostics,} we propose the following concrete research methodologies, focusing on exact training paradigms, loss function adjustments, and dataset requirements.

\subsection{Implicit-Explicit Hybrid Representations}
To bypass the rigid parametric bottleneck imposed by the MHR, future architectures must transition to hybrid representations. We propose integrating 3D Gaussian Splatting \cite{kerbl20233d} or neural signed distance functions (SDFs) directly on top of the explicit parametric mesh, similar to the strategies employed by CHROME \cite{dutta2025chrome}. 

Rather than relying solely on the mesh vertex output $M(\beta, \theta)$, we propose a mathematically hybrid generative framework:
\[
    G = G_{\text{MHR}}(\beta, \theta) + G_{\text{fine}}(\mathcal{G})
\]
where $G_{\text{MHR}}$ generates the low-frequency base mesh to ensure solid topological structure and kinematic accuracy. However, $G_{\text{fine}}$ acts as an additive, high-frequency refinement layer parameterized by a set of dynamic 3D Gaussians $\mathcal{G}$ attached to the mesh surface. 

\textbf{Training Paradigm \& Loss Adjustment:} This hybrid system must be trained via a two-stage curriculum. Stage 1 freezes $G_{\text{MHR}}$ and exclusively trains the Gaussian attributes (opacity, scale, rotation, and spherical harmonics) to capture residual photometric errors. The loss function must be upgraded from standard silhouette or L2 losses to incorporate perceptual metrics that preserve surface texturing:

\[
\begin{aligned}
\mathcal{L}_{\text{hybrid}} &= \lambda_{\text{mesh}} \mathcal{L}_{\text{Chamfer}}(G_{\text{MHR}}, M_{\text{gt}}) \\
&\quad + \lambda_{\text{fine}} \mathcal{L}_{\text{LPIPS}}(R(G), I_{\text{gt}}) + \lambda_{\text{reg}} \mathcal{R}_{\text{density}}(\mathcal{G})
\end{aligned}
\]

where $\mathcal{L}_{\text{LPIPS}}$ ensures high-frequency perceptual matching of the rendered hybrid representation $R(G)$, and $\mathcal{R}_{\text{density}}(\mathcal{G})$ regulates Gaussian densification to prevent artifact accumulation.

\textbf{Dataset Requirements:} Implementing this requires transitioning from weakly-annotated internet images to multi-view, high-resolution clinical capture setups (e.g., photogrammetry rigs synchronized with surface scanners), ensuring the ground truth contains the high-frequency micro-geometry needed to supervise $G_{\text{fine}}$.

\subsection{Domain-Expert Alignment}
We propose that, for a targeted subset of the training data requiring high domain-specific accuracy, annotation be performed exclusively by qualified experts (e.g., board-certified radiologists, anatomists) rather than by generalist annotators. This constitutes a \enquote{Medical-in-the-Loop} pipeline. In this refined alignment phase, the reward function must fundamentally shift from the current generalist objective of \enquote{perceptual quality} to \enquote{pathological fidelity}. 
    
We redefine the Preference Optimization (e.g., DPO \cite{rafailov2023direct}) model. Instead of the standard aesthetic reward:
\[
    r_{\text{general}}(G) = \mathbb{E}_{h \sim \mathcal{H}_{\text{general}}}[\text{aesthetic}(G)]
\]
we engineer a medically-informed deterministic reward:
\[
    r_{\text{medical}}(G, G_{\text{ref}}) = -\mathcal{L}_{\text{EMD}}(G, G_{\text{ref}}) - \lambda_{\text{clinical}} \mathcal{L}_{\text{landmark}}(G, G_{\text{ref}})
\]
where $G_{\text{ref}}$ is the ground-truth medical reconstruction (e.g., from CT/MRI). $\mathcal{L}_{\text{EMD}}$ operates as the Earth Mover's Distance avoiding point-to-point correspondence failures, and $\mathcal{L}_{\text{landmark}}$ penalizes deviations in critical diagnostic landmarks (e.g., ASIS and PSIS in the pelvis).

\subsection{Parametric Injection}
To counteract the 20-component morphological compression, future work must explicitly inject medical shape parameters into the diffusion or regression backbone, adapting the phenotype-inclusive ideology of AnnyBody \cite{bregier2025anny} and SUPR \cite{osman2022supr}. 

By predicting extended spatial shape coefficients ($\beta_{\text{ext}}$) that strictly govern localized volumetric variations (such as distinct parameters governing abdominal distension during pregnancy or localized scoliosis angles), the model logically decouples general \enquote{body type} from \enquote{localized geometry}. The generative mapping expands to:
\[
    G = \mathcal{M}(\beta_{\text{core}} \oplus \beta_{\text{ext}}, \theta) + \Delta(O, \mathbf{c})
\]
where $\beta_{\text{core}}$ retains the standard MHR layout, $\oplus$ denotes concatenation with the newly injected medical parameters, $\theta$ handles poses, and $\Delta(O, \mathbf{c})$ represents a learned graph-convolutional residual deformation conditioned on specific condition labels $\mathbf{c}$ and image observation $O$.

\textbf{Training Paradigm \& Data Requirements:} Training this extended parameter space demands heavily annotated, clinically-segmented datasets where individual subjects are paired with precise medical severity scores (e.g., Cobb angle for scoliosis, specific gestational week for pregnancy). During training, we must inject a disentanglement loss (e.g., an Information Bottleneck or mutual information minimization penalty) to ensure that the network does not inadvertently alter the global $\beta_{\text{core}}$ when adjusting the localized $\beta_{\text{ext}}$.

\section{Conclusion}
In this paper, we dissect SAM 3D Body's consistent difficulty in reconstructing out-of-distribution anthropometric deviations, such as third-trimester pregnancy morphology, severe spinal scoliosis, and geriatric skeletal atrophy. We found that this represents an architectural trade-off favoring high-level perceptual plausibility over strict geometric and topological fidelity rather than a sheer capacity limitation. We isolate three primary mechanisms driving this morphological standardization: First, the MHR's parametric bottleneck rigidly constrains anatomical variance to an overly compressed linear subspace (just 20 core shape coefficients) anchored entirely to standardized body types. Second, DINOv3's inherently invariant tokenization explicitly discards pathologically significant high-frequency variations as algorithmic noise. Finally, annotation-based perceptual alignment actively enforces overarching smoothness priors that penalize and suppress biological irregularities.

This architectural dynamic fundamentally exemplifies the strict perception-distortion compromise in modern computer vision foundation models, explicitly raising significant barriers to their standalone deployment in critical diagnostic frameworks where sub-millimeter anatomical features dictate precise clinical outcomes. Nonetheless, the robustness of SAM 3D Body's latent representations provides a statistical prior for medical simulation. By bridging the representation gap via implicit-explicit hybrid geometric architectures, facilitating targeted fine-tuning on pathological datasets, and pioneering strict Medical-in-the-Loop alignment pipelines, future iterations can successfully extend the utility of these foundational models into clinical diagnostics.

\bibliographystyle{plainnat}
\bibliography{ref}

\end{document}